\documentclass[letterpaper, 10 pt, conference]{ieeeconf}  % Comment this line out if you need a4paper

\IEEEoverridecommandlockouts                              % This command is only needed if 
\usepackage{graphics} % for pdf, bitmapped graphics files
\usepackage{epsfig} % for postscript graphics files
\usepackage{mathptmx} % assumes new font selection scheme installed
\usepackage{times} % assumes new font selection scheme installed
\usepackage{amsmath} % assumes amsmath package installed
\usepackage{amssymb}  % assumes amsmath package installed
\usepackage{multirow}
\usepackage{array}  % Needed for defining a new column type

\newcolumntype{M}{>{\centering\arraybackslash}p{1.5cm}}  % Defining a new column type

\title{\LARGE \bf
Chained Flexible Capsule Endoscope: Unraveling the Conundrum of Size Limitations and Functional Integration for Gastrointestinal Transitivity}

\author{Sishen Yuan$^{1,\dagger}$, Guang Li$^{1,\dagger}$, Baijia Liang$^{1}$, Lailu Li$^{1}$, Qingzhuo Zheng$^{1}$, \\Shuang Song$^{2}$, Zhen Li$^{3}$, Hongliang Ren$^{1*}$%,4
\thanks{The work was supported by Hong Kong Research Grants Council (RGC) NSFC/RGC Joint Research Scheme N\_CUHK420/22; Collaborative Research Fund (CRF C4063-18G, C4026-21G), and General Research Fund (GRF 14203323), Shenzhen-Hong Kong-Macau Technology Research Programme (Type C) Grant 202108233000303 at The Chinese University of Hong Kong (CUHK). (Corresponding author: Hongliang Ren. $\dagger$ indicates contributed equally.) }% <-this % stops a space
\thanks{$^{1}$Sishen Yuan, Guang Li, Baijia Liang, Lailu Li, Qingzhuo Zheng, and Hongliang Ren are with the Department of Electronic Engineering, The Chinese University of Hong Kong, Hong Kong
       }%
\thanks{$^{2}$Shuang Song is with the Harbin Institute of Technology, Shenzhen 518055, China 
       }%
\thanks{$^{3}$Zhen Li is with the Department of Gastroenterology, Laboratory of Translational Gastroenterology, and Robot Engineering Laboratory for Precise Diagnosis and Therapyof GI Tumor, Qilu Hospital of Shandong University, Cheeloo College of Medicine, Jinan, Shandong, China}
}

\begin{document}

\maketitle
\thispagestyle{empty}
\pagestyle{empty}

%%%%%%%%%%%%%%%%%%%%%%%%%%%%%%%%%%%%%%%%%%%%%%%%%%%%%%%%%%%%%%%%%%%%%%%%%%%%%%%%
\begin{abstract}

Capsule endoscopes, predominantly serving diagnostic functions, provide lucid internal imagery but are devoid of surgical or therapeutic capabilities. Consequently, despite lesion detection, physicians frequently resort to traditional endoscopic or open surgical procedures for treatment, resulting in more complex, potentially risky interventions. To surmount these limitations, this study introduces a chained flexible capsule endoscope (FCE) design concept, specifically conceived to navigate the inherent volume constraints of capsule endoscopes whilst augmenting their therapeutic functionalities. The FCE's distinctive flexibility originates from a conventional rotating joint design and the incision pattern in the flexible material. In vitro experiments validated the passive navigation ability of the FCE in rugged intestinal tracts. Further, the FCE demonstrates consistent reptile-like peristalsis under the influence of an external magnetic field, and possesses the capability for film expansion and disintegration under high-frequency electromagnetic stimulation. These findings illuminate a promising path toward amplifying the therapeutic capacities of capsule endoscopes without necessitating a size compromise.

\end{abstract}

%%%%%%%%%%%%%%%%%%%%%%%%%%%%%%%%%%%%%%%%%%%%%%%%%%%%%%%%%%%%%%%%%%%%%%%%%%%%%%%%
\section{Introduction}

Capsule endoscopy facilitates non-invasive inspection of the gastrointestinal tract \cite{iddan2000wireless}. Since its introduction into clinical practice in 2001, it has garnered widespread adoption globally. Capsule endoscopy exhibits notable superiority in diagnosing gastrointestinal maladies, but often, diagnosis is merely an initial step towards resolving the issue \cite{wang2013wireless}.

Upon identification of a lesion or anomaly, subsequent treatment or intervention is typically necessitated, frequently requiring more intricate surgical or therapeutic procedures. These encompass biological tissue extraction, drug delivery, haemostasis, among other interventions. However, the current suite of capsule endoscopes is predominantly deployed for diagnostic, as opposed to therapeutic, objectives. They yield lucid internal imagery but are bereft of capabilities to execute surgical or therapeutic manoeuvres \cite{koulaouzidis2023current}.
Consequently, despite lesion detection by the capsule endoscope, physicians often resort to traditional endoscopic or open surgical procedures for treatment. This necessitates a more complex and risky treatment regimen, potentially prolonging the patient's recovery. Therefore, there is an unmet need to enhance the therapeutic capabilities of capsule endoscopy.

\begin{figure*}
    \centering
    \includegraphics[width=0.9\linewidth]{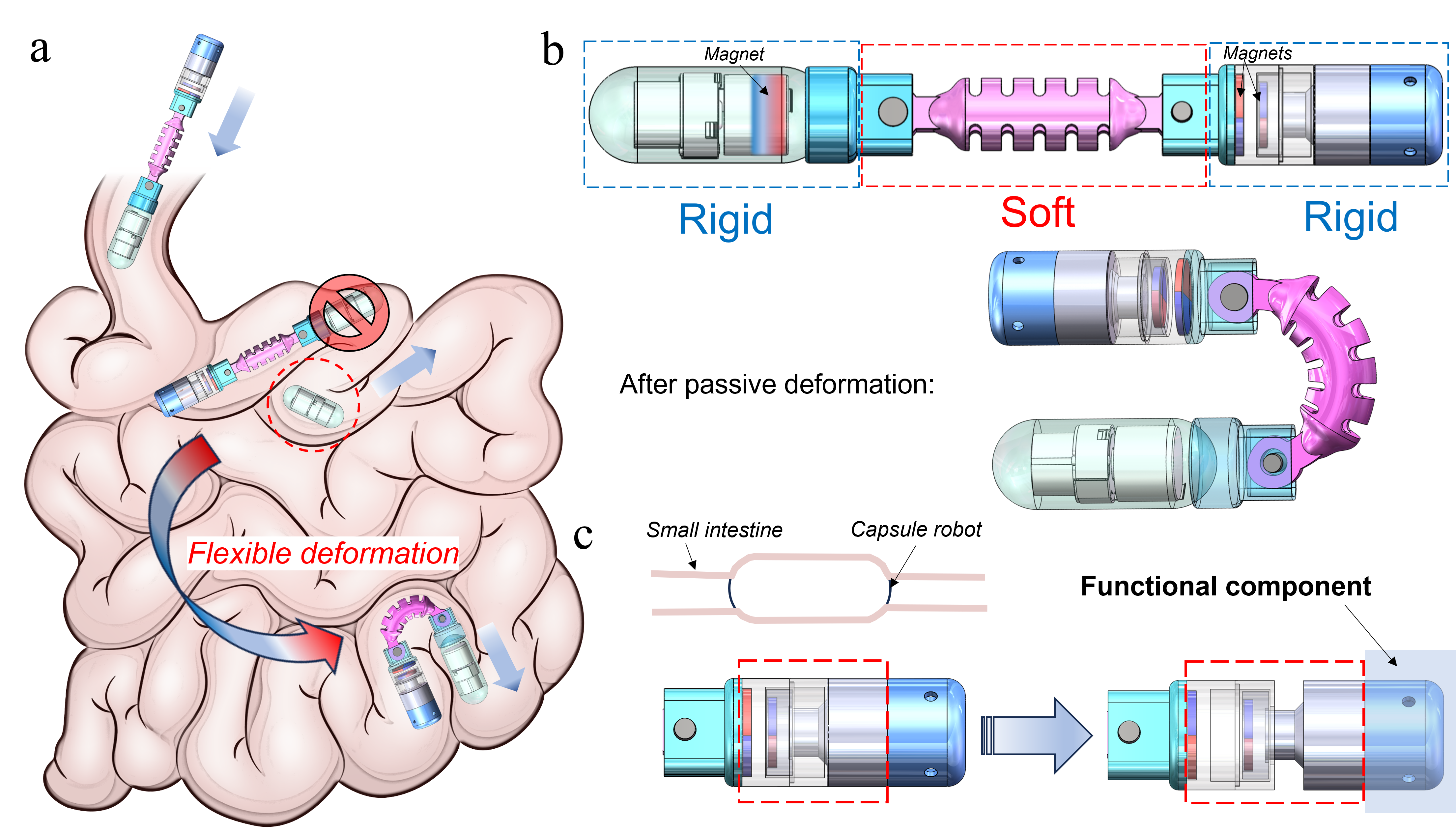}
    \caption{Design and functionality of the proposed flexible capsule endoscope. (a) Illustration of the challenges faced by longer capsule endoscopes during transit through the narrow, elongated lumen of the small intestine, where their smooth passage is impeded. (b) Schematic representation of the "rigid-soft-rigid" configuration of our flexible capsule endoscope. The first 'rigid' segment is modelled after existing commercial endoscopes, with the 'soft' segment enabling passive deformation under intestinal compression stresses and peristaltic forces, facilitating adaptation to the zigzag environment of the small intestine. (c) The additional rigid section at the tail end of the capsule is designed to induce a worm-like crawling propulsion motion via the interaction of a pair of internal radial magnets under an external magnetic field, and to integrate functional components. The figure also depicts the expansion of the capsule endoscope in response to high-frequency electromagnetic heating, leading to the expansion of an Ecoflex film. This feature suggests potential applications for anchoring, drug delivery, sampling, and scaffold deployment.}
\end{figure*}

Numerous contemporary studies concentrate on functionalizing capsule endoscopes, primarily for anchoring, biopsy, and drug delivery \cite{chen2022magnetically}. Anchoring is generally achieved through electrothermal actuation of memory alloy wires or wireless magnetic control. Implementing these mechanisms necessitates either an enlargement of the capsule endoscope's radial dimensions or an increase in its axial rigidity scale \cite{zhou2016magnetically,song2021integrated,wang2020ultrasound}. Current dimensions do not appear to account for the volume occupation of the endoscope's actual vision module. This paradox is similarly observed in biopsy-capable capsule robots \cite{hoang2020robotic,ye2022design,gerbonimagnetic}.

In response to these challenges, a collaborative strategy involving modular capsule robots has been suggested for capsule robot volume expansion. During the examination process, a capsule robot equipped with an image acquisition function is initially used as the primary functional module. Subsequently, according to the patient's individual needs, an additional functional module with corresponding capabilities is ingested. These multiple modules have the potential to assemble in vivo and collectively execute medical tasks \cite{nagy2008experimental,guo2017development,guo2018design}. However, the in vivo assembly and control of these modules remain formidable challenges. Additionally, their suitability for smooth intestinal passage and potential perforation risks have not been fully explored.

This study presents a design concept for flexible capsule endoscopes (FCEs), devised to reconcile the inherent volume constraints of capsule endoscopes with the requirement for comprehensive functionalization. The flexibility of the FCE stems from the traditional rotating joint design and the incision pattern within the flexible material. We substantiated its feasibility for passive navigation through a rugged intestinal tract via in vitro intestinal passage experiments. Furthermore, we experimentally confirmed the consistent reptile-like peristaltic motion of the FCE's tail propulsion mechanism and the entire body under the influence of an external magnetic field. Ultimately, we demonstrate that the FCE can achieve film expansion and overall disintegration under high-frequency electromagnetic stimulation.
\section{System Overview}

The gastrointestinal (GI) tract region most rigorously navigated by capsule endoscopes is the small intestine, notable for its slender form ($\sim$8 m) \cite{wang2006physiological}, narrow cross-section, and thin intestinal wall. The passage capability of currently available commercial endoscopes through this area is well-documented, albeit with isolated incidences of hysteresis. Yet, an increase in the length of capsule endoscopes evidently presents a challenge to their smooth transit through this lumen, as illustrated in Figure 1a.

The FCE adopts a "rigid-soft-rigid" configuration (Figure 1b). The rigid head section, typically around 12 mm in diameter and 27 mm in length, aligns with the dimensions of standard commercial endoscopes. The central soft section, pivotal in extending the length of the capsule endoscope, facilitates passive deformation under intestinal compressive stresses and peristaltic forces. This design draws inspiration from the mechanics intrinsic to traditional rotary joints and flexible surgical robots. Consequently, an additional rigid section can be affixed at the tail end, a feature designed to enhance the diverse functionalization of the FCE.

As the FCE navigates the small intestine, it frequently encounters narrowed regions, enveloped by the inner intestinal wall. To facilitate the capsule endoscope's progress within the small intestine, we have engineered a pair of radial magnets at the tail end (Fig. 1c).
Under the influence of their internal magnetic potential energy, they naturally adopt a closed state, but are compelled to separate by the exertion of an external magnetic field. As a result, a worm-like bionic reciprocal motion in the axial direction can be realized. This offers propulsion to the head's rigid part, bolstered by the reactionary force of the intestinal wall.

Furthermore, weak acid and alkali solutions are integrated into the tail's functional components, segregated by hot-melt adhesive and coated round iron sheet. Under high-frequency electromagnetic heating, the hot melt transitions into a fluid state, instigating an acid-base reaction that generates carbon dioxide. This elevates the internal pressure, culminating in the expansion of the Ecoflex film. This innovative feature paves the way for future capsule endoscopes, opening up possibilities for soft anchoring, drug delivery, sampling, and stent deployment.

\section{FCE connector design and evaluation}

\begin{figure}
    \centering
    \includegraphics[width=0.9\linewidth]{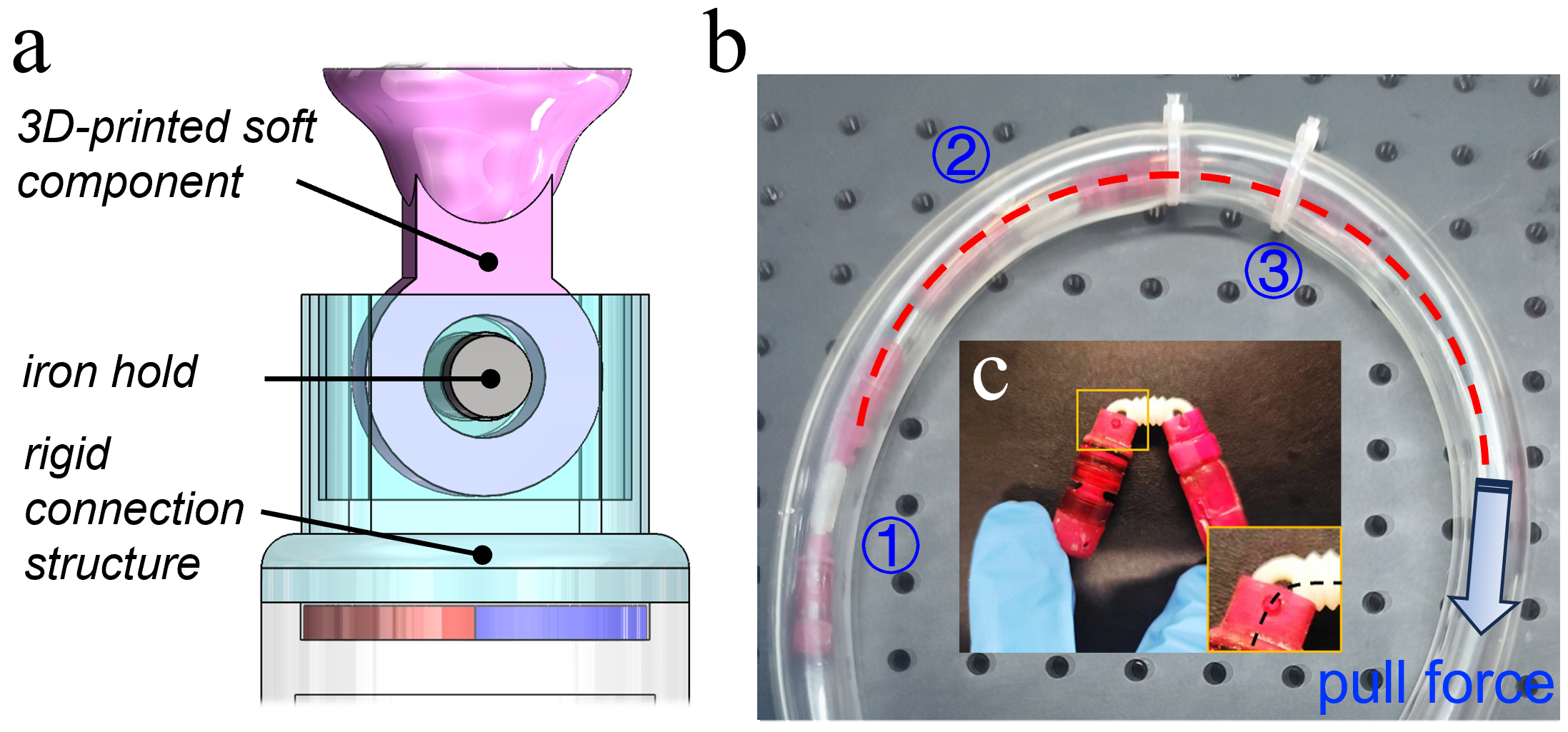}
    \caption{Schematic of the rotary joint in a flexible capsule, constructed with a custom connection mechanism, 3D-printed malleable ring, and iron pins. (b) "Passability" metric used for performance evaluation, where a tethered wire in a curved plastic hose replicates small intestine conditions. Results are summarized in Table 1.(c) Manual bending of the flexible capsule and measurement of extreme bending angle. Despite a stiffness of 70\%, the softer, non-hollowed-out components do not cause appreciable deformation, with bending facilitated by the rotary joints and ring deformation. Component lengthening from 18mm to 24mm did not notably increase the bending angle. These findings may inform future design optimizations of flexible capsules.}
\end{figure}

% \usepackage{multirow}
\begin{table}[]
\caption{Examination results of material length and stiffness on the rotary joint performance in a flexible capsule robot.}
\label{tab1}
\begin{tabular}{|c|c|c|c|}
\hline
Length (mm)         & Stiffness factor (\%) & Passability & \multicolumn{1}{l|}{\begin{tabular}[c]{@{}l@{}}Extreme bending \\    angle ( °)\end{tabular}} \\ \hline
\multirow{2}{*}{12} & 70                   & Yes         & 160                                                                                           \\ \cline{2-4} 
                    & 85                   & No          & 145                                                                                           \\ \hline
\multirow{2}{*}{18} & 70                   & Yes         & 180                                                                                           \\ \cline{2-4} 
                    & 85                   & No          & 60                                                                                            \\ \hline
\multirow{2}{*}{24} & 70                   & Yes         & 180                                                                                           \\ \cline{2-4} 
                    & 85                   & No          & 60                                                                                            \\ \hline
\end{tabular}
\end{table}

The configuration of the connection mechanism within an FCE is integral to its performance. We have incorporated a classical rotary joint design (Fig. 2a) to enhance passive bending flexibility. The joint comprises a bespoke connection structure and a 3D-printed ring, fabricated from an elastic material, which are interconnected via an iron hold.
The soft connector is fabricated using a Stratasys J826 3D printer (Stratasys, Eden Prairie, MN, USA) with Agilus30 Clear (Stratasys) serving as the print material.
Subsequently, we adjusted the length and stiffness factor of the material to empirically investigate the direct influence of these physical quantities on the results. A succinct encapsulation of the findings is presented in Table \ref{tab1}.
To evaluate whether this design facilitates an increase in the rigid length of the capsule endoscope, we introduced a metric termed "passability". A PVC clear plastic hose was employed to fabricate a curved environment with a bending angle of $\sim$300°, thereby replicating the extreme deformation angles of the small bowel in vivo. A wire attached to the flexible capsule simulated the peristaltic forces exerted by the small intestine (Fig. 2b).
Furthermore, manual bending of the flexible capsule was conducted (Fig. 2c), followed by measurement of the ultimate bending angle. Observations indicated that even a 70\% stiffness does not induce significant deformation as the softer components are not hollowed out. Rather, the rotary joint's rotation and the ring's deformation facilitated bending. Notably, extending the part from 18 to 24 mm does not appreciably augment the ultimate bending angle. These insights suggest that additional modifications to the primary section of the soft component may be required to optimize the bending performance of FCE.

\begin{figure}
    \centering
    \includegraphics[width=0.9\linewidth]{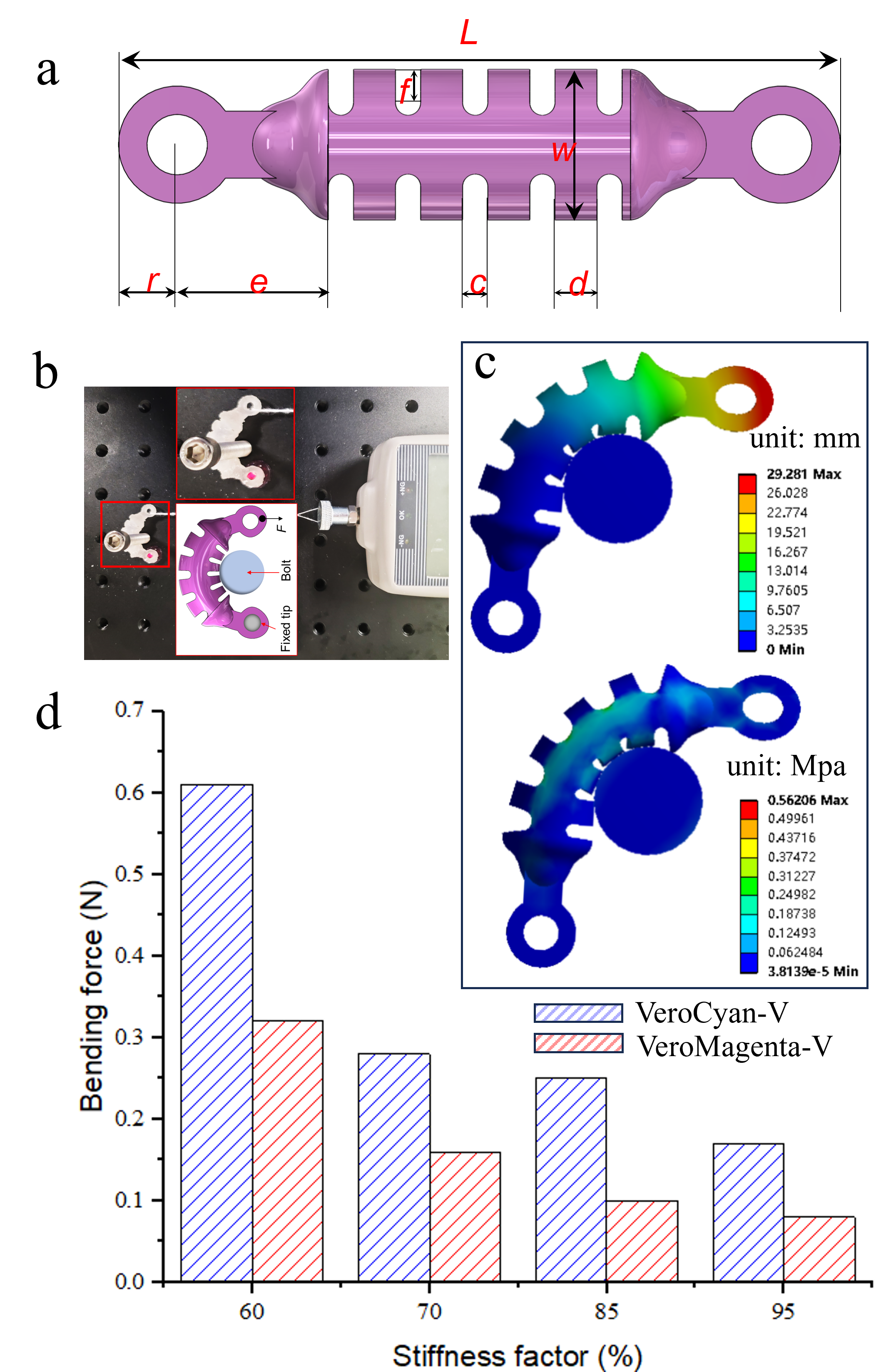}
    \caption{Analysis and Modelling of 'Soft' Structure. (a) Shows parametric representation of the structure. (b) Explores material property impact on bending deformation. (c) Presents ANSYS simulation results aligning with experimental data. (d) Indicates lower bending force with higher stiffness, favouring VeroMagenta-V over VeroCyan-V as a connecting material.}
\end{figure}

\begin{figure}
    \centering
    \includegraphics[width=0.9\linewidth]{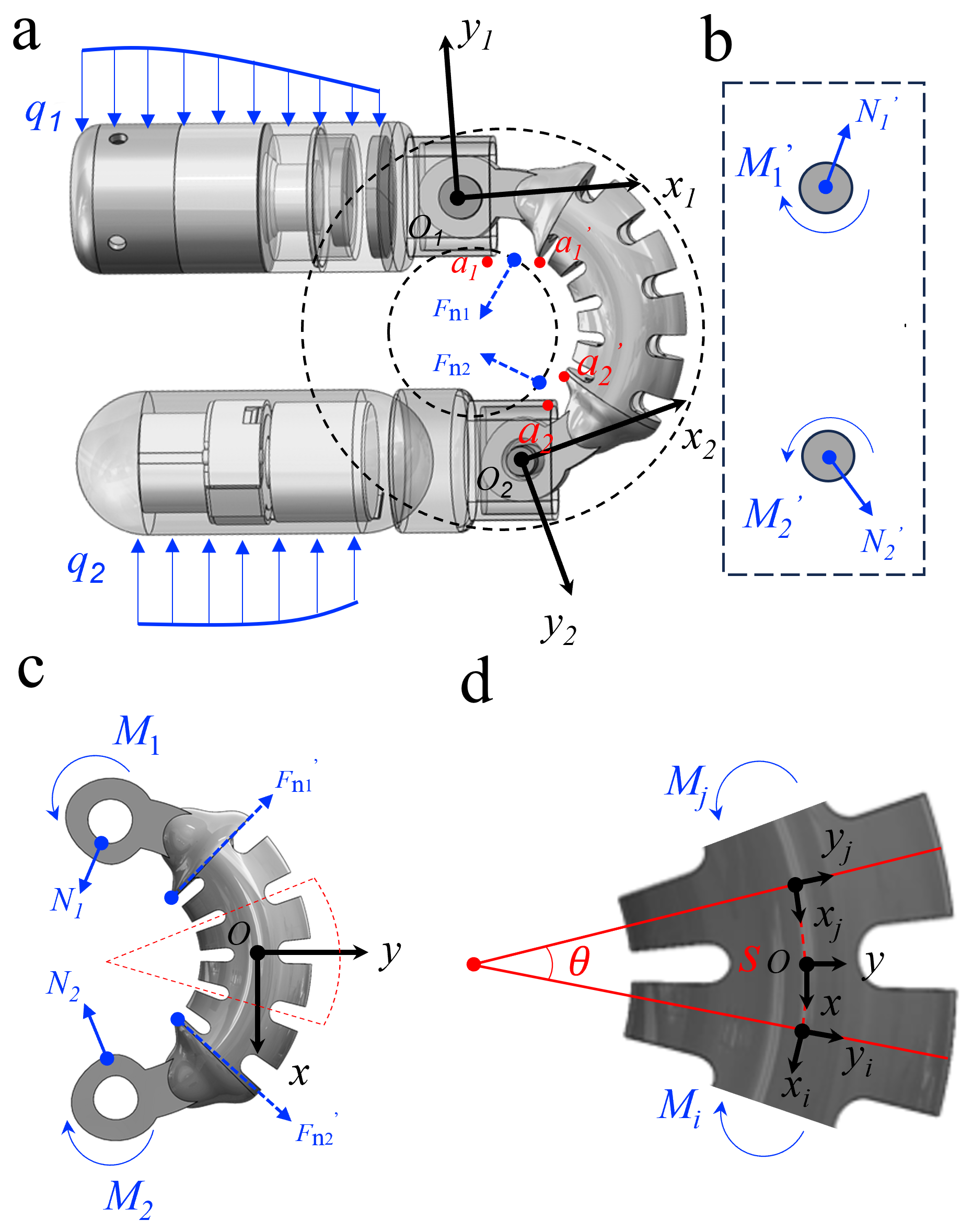}
    \caption{Deformation Analysis of FCE. (a) Shows force analysis with stresses $q_{1}$, $q_{2}$ and reaction forces $\textbf{F}{n1}$ and $\textbf{F}{n2}$. Deforming areas $a_{1}^{'}$ and $a_{2}^{'}$ enable wider bending. (b) Displays passive deformation via iron hold forces and reaction forces $\textbf{F}{n1}^{'}$ and $\textbf{F}{n2}^{'}$ (c). (d) Examines soft section deformation under section moments $\textbf{M}{i}$ and $\textbf{M}{j}$. The moment-curvature relationship is derived from the Euler-Bernoulli beam theory.}
\end{figure}

\begin{figure}
    \centering
    \includegraphics[width=0.9\linewidth]{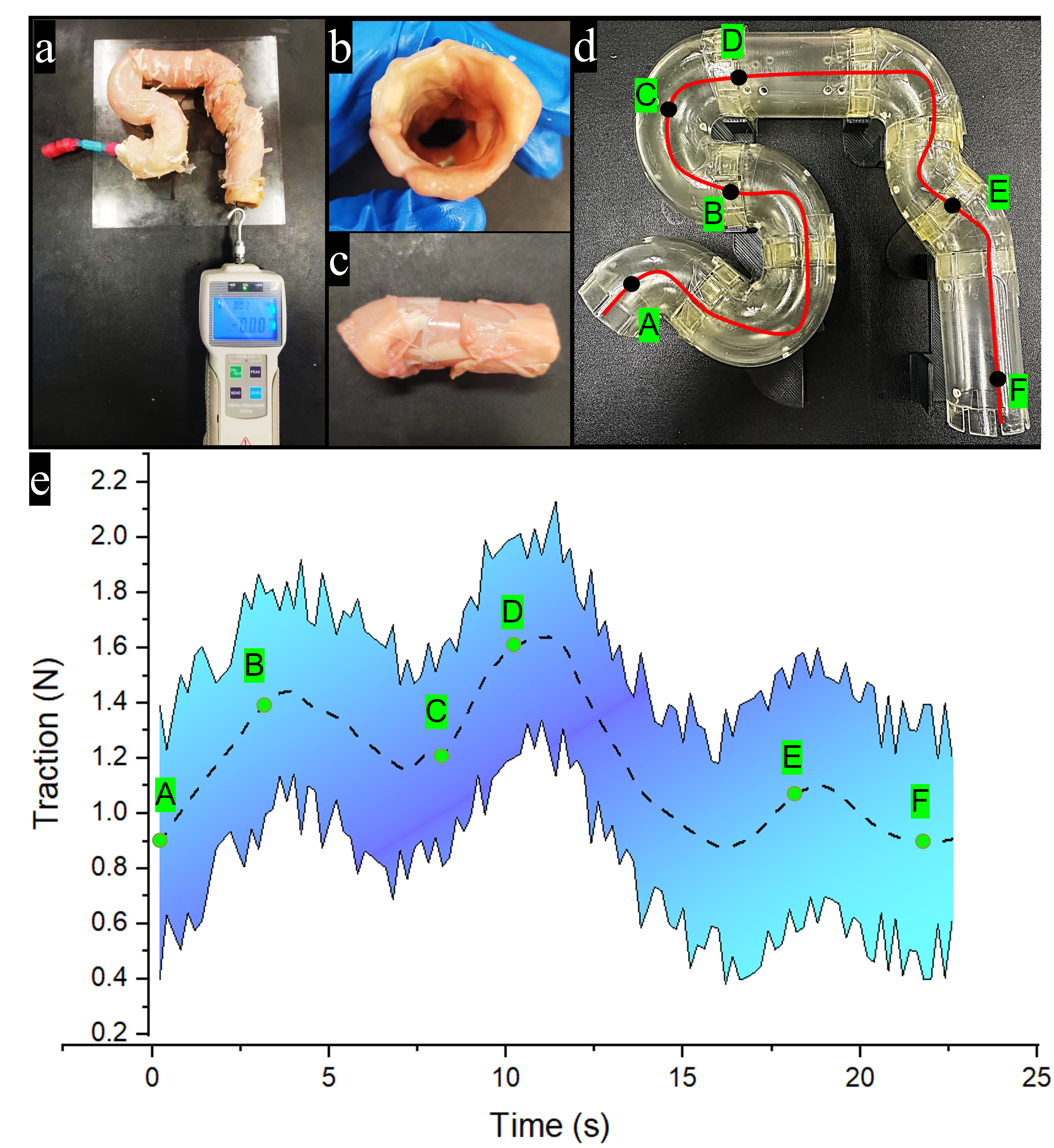}
    \caption{(a) In vitro intestinal passage experiment to assess the passive transport capability of the FCE. (b) Model's interior, constructed from pig small intestine, emulates a true intestinal wall. (c) Depiction of one section of the model. (d) Model's rigid, zigzag-patterned support made of acrylic. (e) Force associated with the passive transit of the FCE along the path, demonstrating an elevated tensile force requirement at large corners and small curvatures, such as points B and D, yet confirming the FCE's effective adaptability to such conditions.}
\end{figure}

Incision tube structures, also referred to as torsos, comprise incision tubes engineered to diminish rigidity and permit bending when externally actuated via wires. Lateral incision tube structures, fabricated from superelastic materials, are gaining prominence in the field of surgical robotics. Their seamless integration with surgical instruments and provision of enhanced maneuverability for surgeons is highly valued \cite{legrand2021large}. We were inspired by their ability to attain large bending angles and tight bending radii for the design of the primary structure of the soft segment, as depicted in Fig. 3. This makes it particularly adept at maneuvering in narrow lumens.

Figure 3a illustrates the "soft" structure's parametric representation, as summarized in Table. \ref{tab2}.
Figure 3b elucidates the impact of printed material properties on the bending of a soft segment. One end of the segment is secured, while the other is affixed to a force transducer using screws to constrain displacement and engender the desired bending deformation. 
An analogous simulation was performed utilizing \cite{sun2022larg}, where Figure 3c presents the resultant, converged displacement and stress distributions. These outcomes exhibit a satisfactory congruence with the empirical data. Furthermore, the experimental findings, as depicted in Fig. 3d, reveal that the required bending force inversely correlates with the stiffness factor. This suggests that VeroMagenta-V exhibits superior suitability as a connecting segment compared to VeroCyan-V \cite{kusaka2015initial}.

\begin{table}[]
\centering
\caption{Parameters of the Prototype's 'Soft' Section (Unit: mm)}
\label{tab2}
\begin{tabular}{|l|l|l|l|l|l|l|}
\hline
L  & f & w & r   & e & c   & d   \\ \hline
43 & 2 & 9 & 3.5 & 9 & 1.5 & 2.5 \\ \hline
\end{tabular}
\end{table}

Figure 4a details the force analysis of the FCE under passive deformation. Here, $q_{1}$ and $q_{2}$ represent the stresses applied to the rigid section of the intestinal wall. The points of contact, $a_{1}$ and $a_{2}$, between the rigid connection structure and the 3D-printed soft component at the FCE's maximum bending, generate respective reaction forces $\textbf{F}_{n1}$ and $\textbf{F}_{n2}$, thereby constraining further deformation. As $a_{1}^{'}$ ($a_{2}^{'}$) on the soft section undergoes deformation, it offers the FCE a broader bending angle across its body. The passive deformation of the soft section is facilitated by the torque and compression forces transmitted by the iron hold (Fig. 4b), along with the reaction forces $\textbf{F}_{n1}^{'}$ and $\textbf{F}_{n2}^{'}$, in the ultimate bending state (Fig. 4c).
The observed deformation within the soft section is characterized by the bending of notched cells under the influence of section moments  $\textbf{M}_{i}$ and $\textbf{M}_{j}$ (Fig. 4d). Extending the principles of Euler-Bernoulli beam theory to encompass constant-curvature beams, we establish a moment-curvature relationship applicable to a generally curved beam, as expressed herein\cite{zeng2021modeling}:
\begin{align}
\setlength{\abovedisplayskip}{3pt}
\setlength{\belowdisplayskip}{3pt}
\small
\frac{d\theta }{ds} = \frac{1}{R} - \frac{\textbf{M}_{i}+\textbf{M}_{j}}{EI} 
\label{mag}
\end{align}
where $\frac{d\theta }{ds}$ is the curvature at $s$ and $s$ is the curvilinear coordinate of the soft section. Parameters $R$, $E$, and $I$ denote the initial radius of curvature, Young's modulus, and the second moment of the equivalent area, respectively.

To assess the passability of FCE, an in vitro intestinal passage experiment was conducted (Fig. 5a). Here, FCE was maneuvered through an in vitro model via a rope tether connected to a force transducer. The model's interior was constructed using pig small intestine to mimic a genuine intestinal wall (Fig. 5b), with a section's appearance displayed in Fig. 5c. The model's rigid support, crafted from acrylic, features a zigzag design (Fig. 5d). Fig. 5e illustrates the force involved in the passive movement of the FCE along the path, highlighting an increase in the necessary tensile force for large corners and small curvatures, such as points B and D. Nevertheless, passage is feasible, affirming the FCE's superior adaptability to such environments.

\section{Integrated Functions}
\subsection{Active Locomotion}

\begin{figure}
    \centering
    \includegraphics[width=1.0\linewidth]{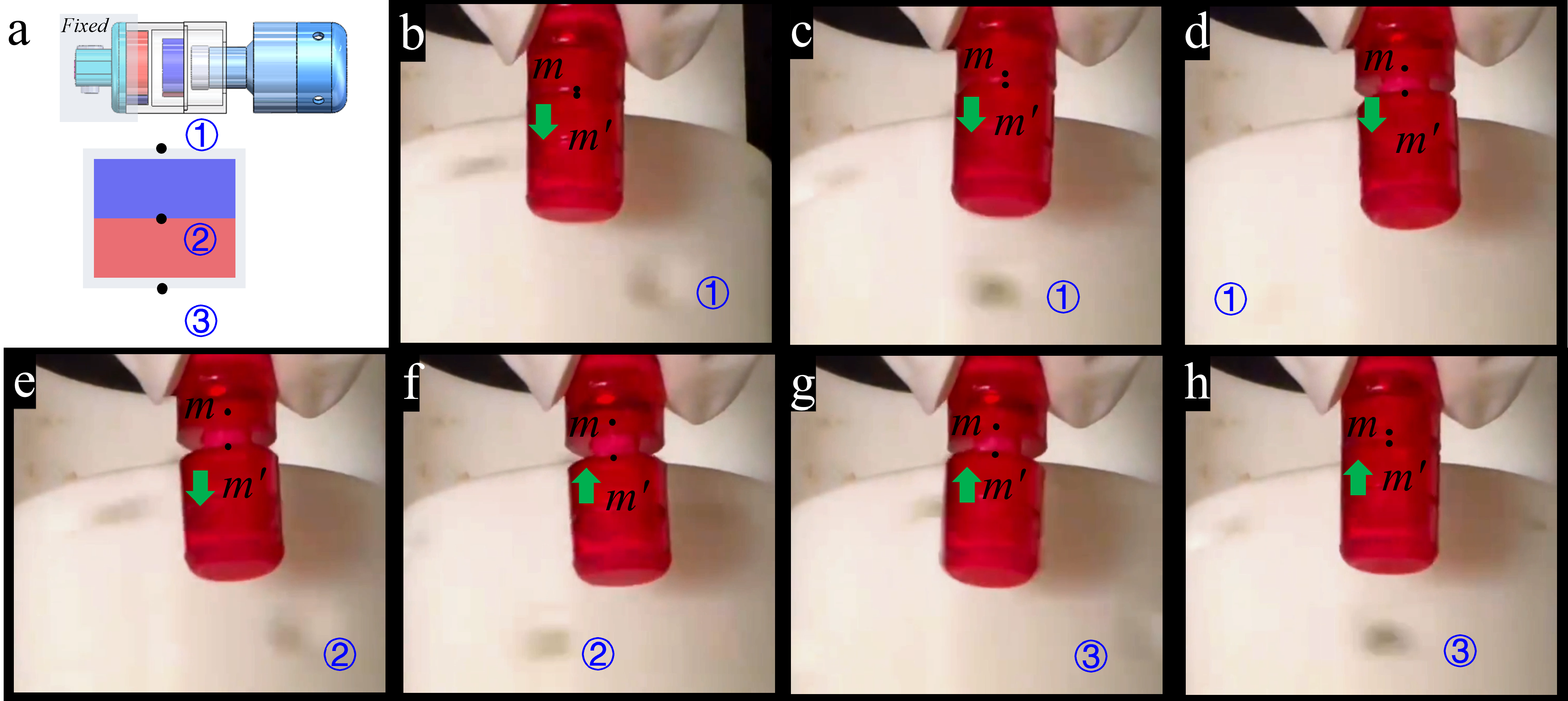}
    \caption{(a) Schematic of the experiment where the FCE's head end is fixed and external drive magnets rotate at 3Hz. (b-h) Sequence showing the combined rotational and linear motion of the FCE tail, beneficial for forward propulsion in confined environments.}
\end{figure}

We confirmed the FCE's capability for passive intestinal delivery, and subsequently conducted experimental verification of its ability to be remotely actuated under the influence of an external magnetic drive system.
Initially, we verified that the internal radial magnet pair could convert the input rotating magnetic field signal into axial displacement. Fig. 6a presents a schematic, where the head end of the FCE is fixed, and the external drive magnets are affixed to a torque motor rotating at 3Hz, with marker points designated on the drive housing. Fig. 6 (b-h) depicts a motion sequence in which the FCE tail can generate combined rotational and linear motion. This feature may prove beneficial in confined environments, where the tail facilitates forward propulsion by interacting with direct tissue contacts.

Building on this, we employed the in vitro intestinal model described in Fig. 5(d), boasting a diameter of approximately 20 mm, to authenticate the active penetration capabilities of the FCEs within a tortuous, large cavity. The relative positions of the in vitro gut model and the driving magnet were manually adjusted, and the FCE exhibited a sensitive response (Fig. 7(a-f)). When the FCE encountered a challenging zigzag narrow space (Fig. 7(g,h)), the tail mechanism executed a propulsive function to aid the FCE's onward progression as the head's forward travel faced constraints.

Prior experiments indicated a vigorous movement of FCEs that could prove unsuitable for passage through the thin-walled small intestine, potentially posing risks such as perforation. Drawing on methods from previous research, we validated the approach in a porcine small intestine. Employing a wire, a live knot was attached to a segment of the small intestine to simulate an intestinal narrowing. The FCE navigated through the small intestine with gentle motion and successfully traversed the live knot opening without difficulty (Fig. 8).

\begin{figure}
    \centering
    \includegraphics[width=1.0\linewidth]{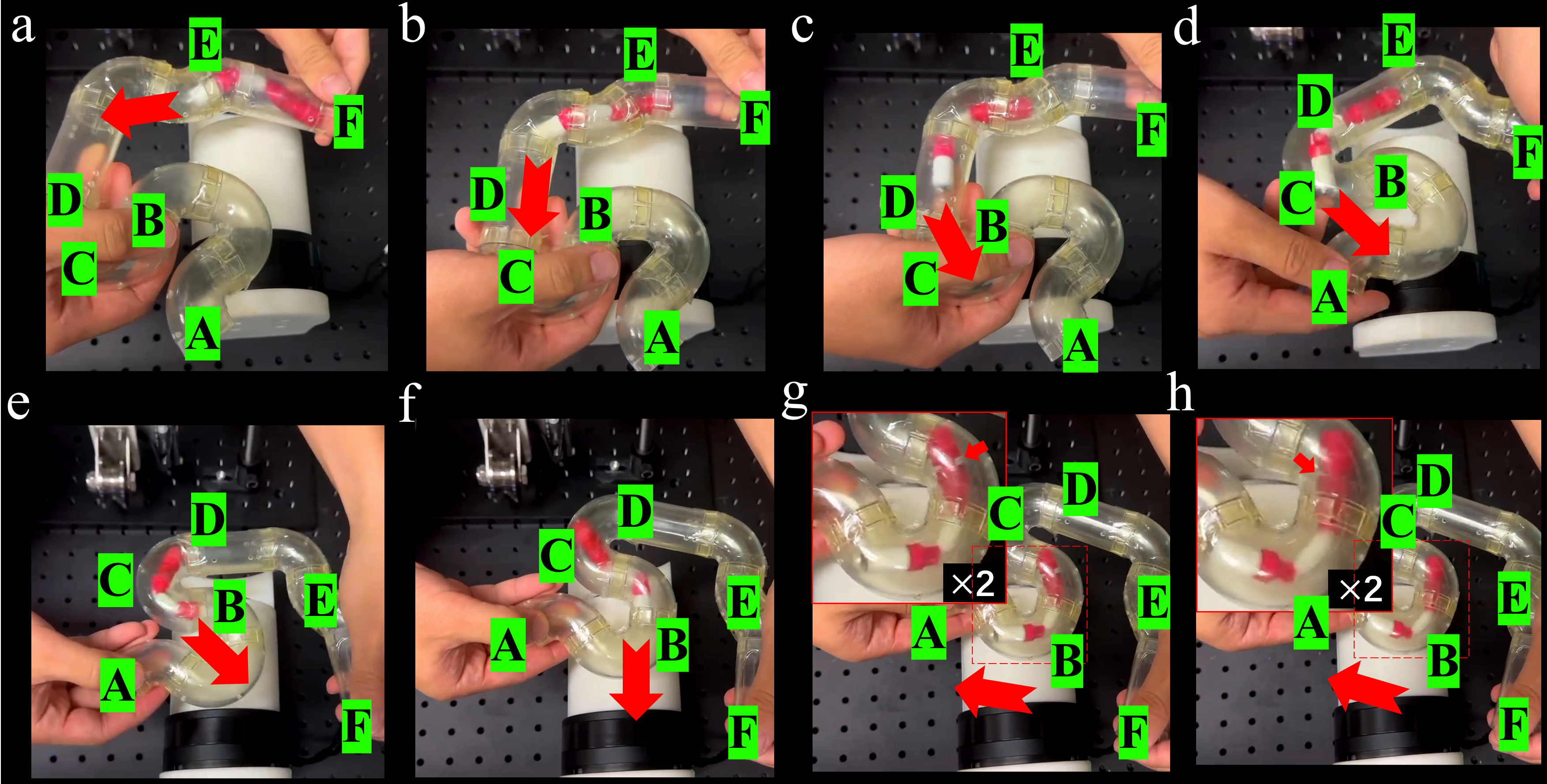}
    \caption{(a-f) FCE's response in the in vitro intestinal model, showcasing its penetration capabilities. (g-h) FCE's navigation in challenging zigzag spaces, with the tail aiding forward progression when limited.}
\end{figure}

\begin{figure}
    \centering
    \includegraphics[width=0.85\linewidth]{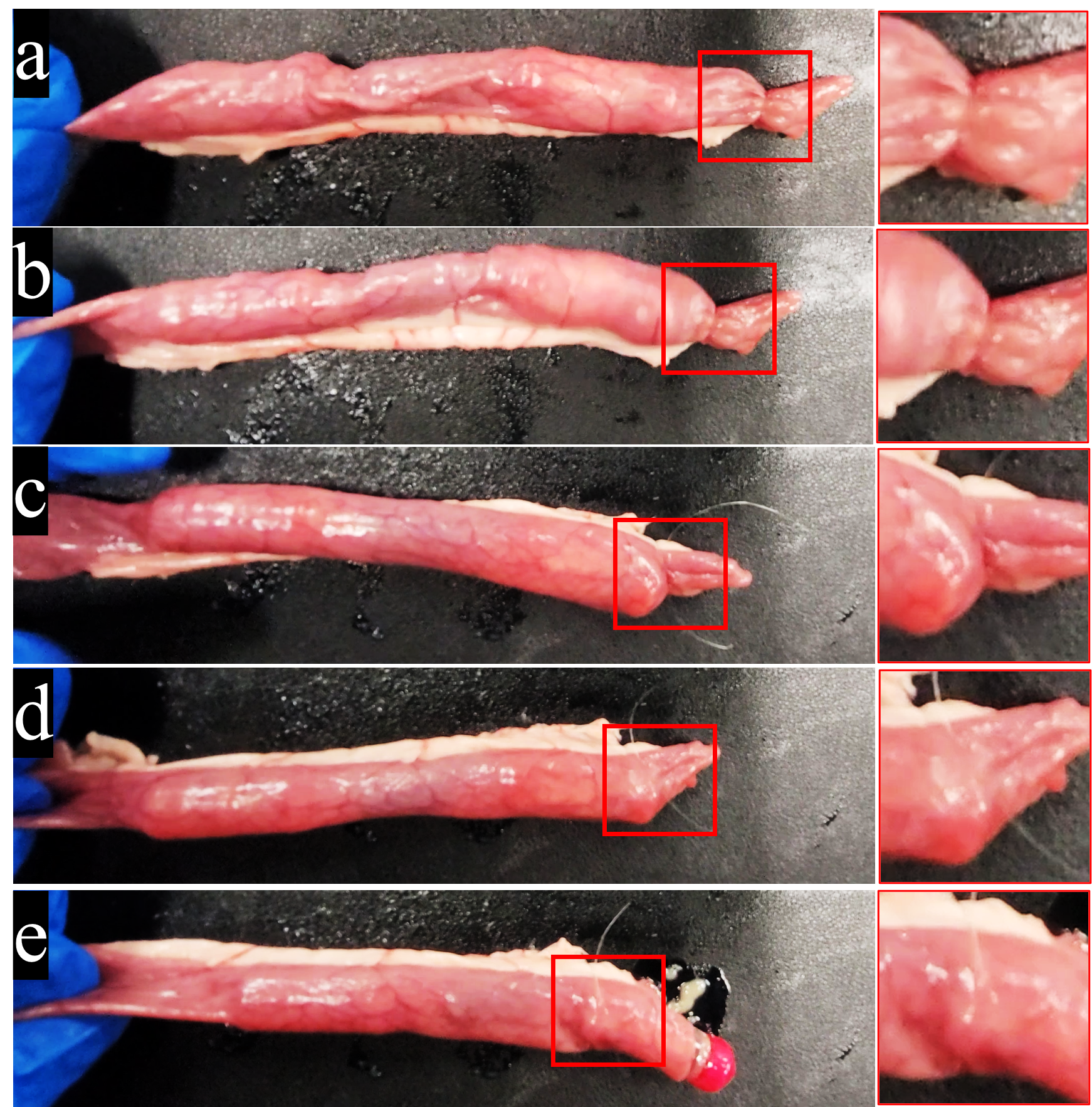}
    \caption{Validation of FCE's movement in a porcine small intestine, addressing previous concerns of potential perforation due to vigorous motion. A live knot, simulating intestinal narrowing, was attached via a wire. The figure illustrates the FCE's gentle navigation and successful traversal through the live knot opening.}
\end{figure}

\subsection{Expansion and Separation}

The FCE allows for an increase in capsule endoscope size, mitigating concerns of retention in the gastrointestinal (GI) tract. To illustrate this FCE advantage, we developed a thermal trigger mechanism to control the acid-base reaction within the rigid structure of the FCE tail, encased in an ecoflex film. The thermal triggering mechanism relies on a hot melt adhesive, which melts at approximately 50°C \cite{yang2023reversible}. Installed in a round iron plate as a valve in its hot melt state, it is heated using a high-frequency electromagnetic field. Upon transformation to a liquid state, the valve opens, and the acid/base reaction generates carbon dioxide and water, enabling the film's expansion. This membrane expansion, a basic function, can be further utilized for applications such as intestinal anchoring, biopsy of the intestinal wall and drug administration in combination with microneedles, and potentially stent deployment in narrow lumens (Fig. 9).

\begin{figure}
    \centering
    \includegraphics[width=1.0\linewidth]{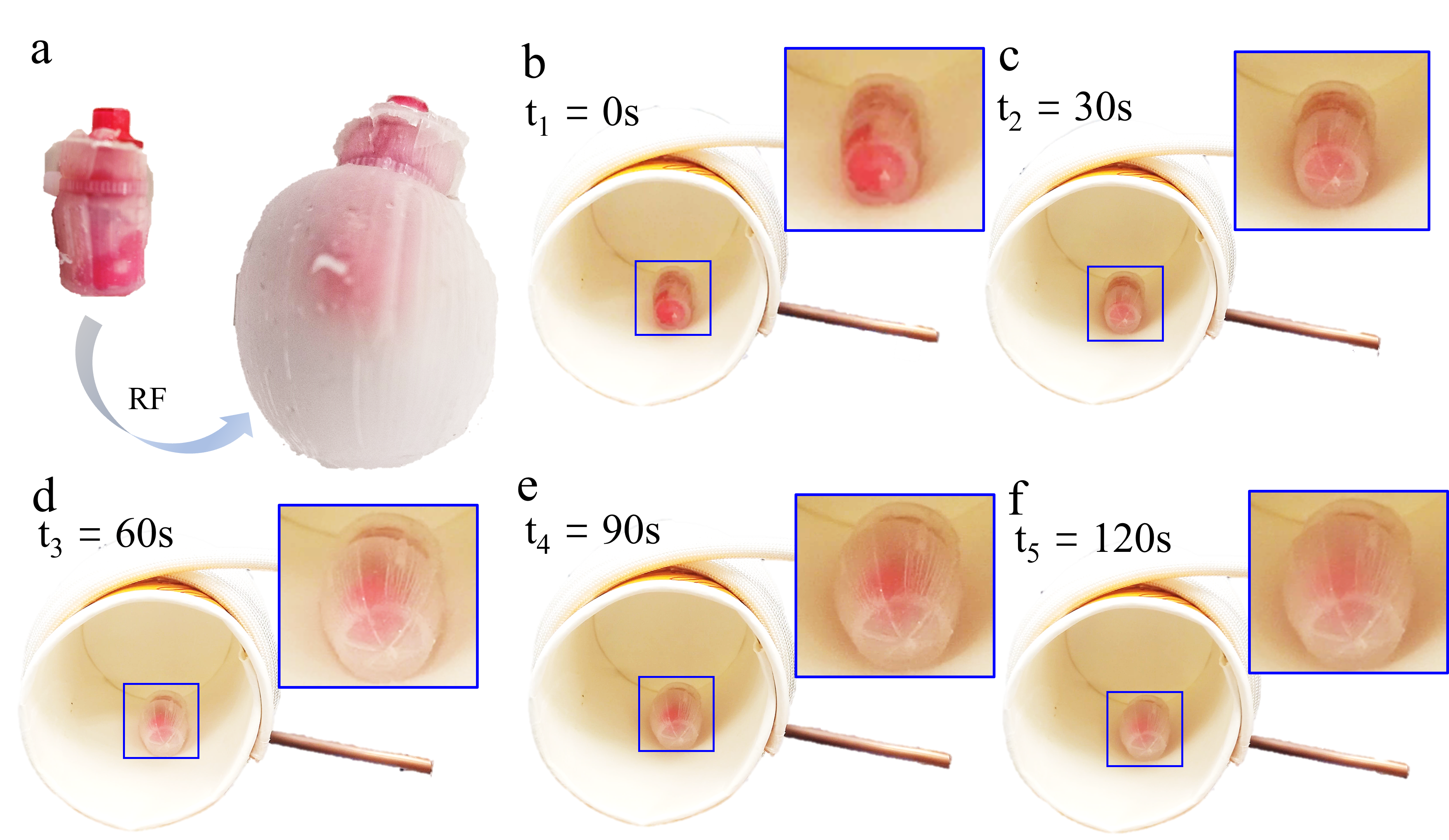}
    \caption{Demonstration of the FCE's thermal trigger mechanism. The mechanism, reliant on a hot melt adhesive, triggers an acid/base reaction upon heating, leading to film expansion. This can be used for applications like intestinal anchoring, wall biopsy, drug administration, and potential stent deployment.}
\end{figure}

In line with the aforementioned principle, we've substituted a portion of the FCE's connecting ring with a hot melt adhesive. The thermal effect of the iron pin in the high-frequency electromagnetic field allows for the controlled disconnection of the connecting ring. Consequently, the FCE can disassemble and be naturally expelled from the body upon completion of its function (Fig. 10).

\begin{figure}
    \centering
    \includegraphics[width= 0.8\linewidth]{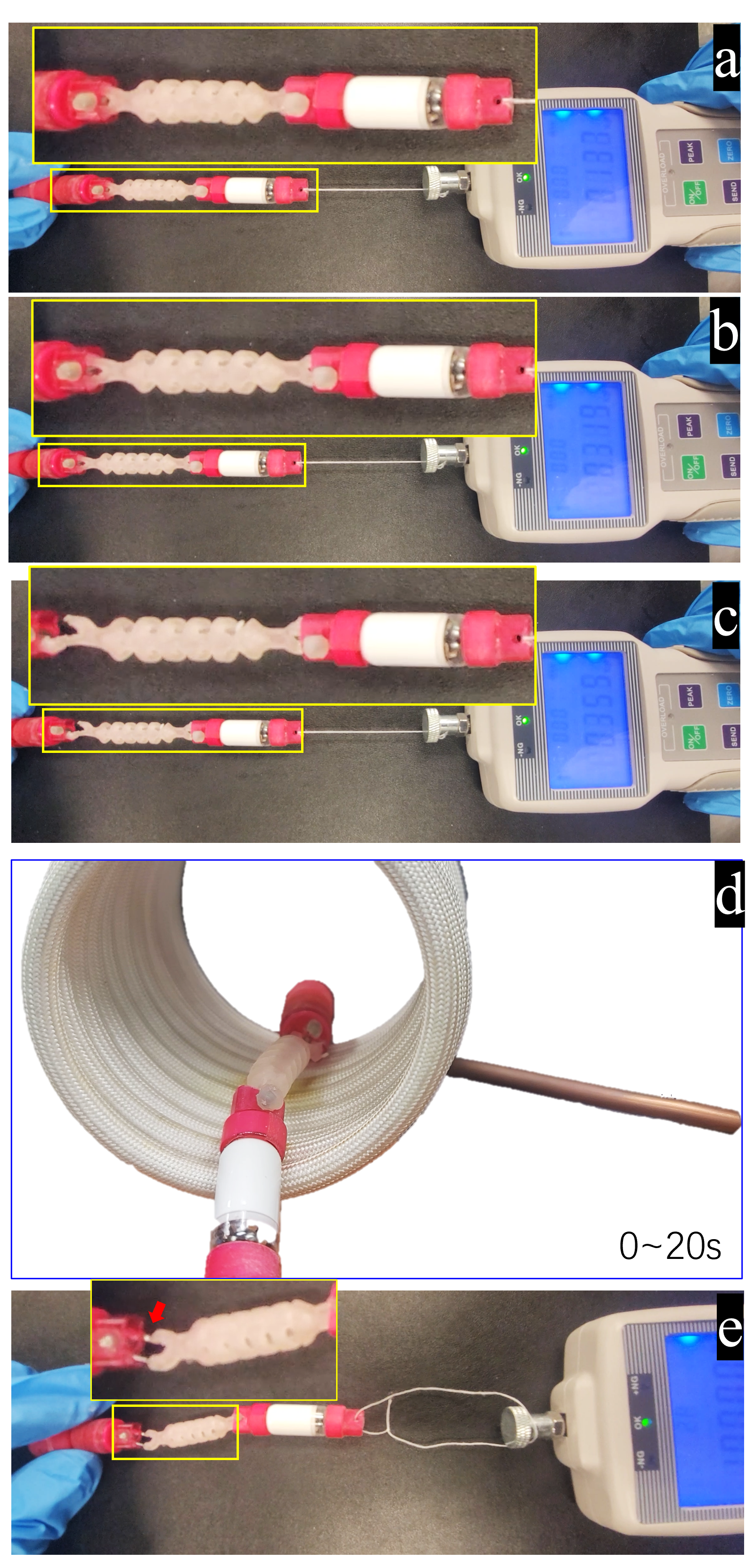}
    \caption{Depiction of FCE's modified connecting ring with hot melt adhesive. The thermal effect allows controlled ring disconnection, facilitating FCE disassembly and natural expulsion post-function.}
\end{figure}

\section{Discussion and Conclusion}

%The small intestine, with its elongated, narrow structure and thin walls, is the region of the gastrointestinal (GI) tract most safely traversed by capsule endoscopes. Existing commercial endoscopes demonstrate proficient transit through this region, with infrequent retention incidents. However, as the length of the capsule endoscopes augments, challenges to their smooth passage through this lumen emerge.

Our proposed flexible capsule endoscope incorporates a "rigid-soft-rigid" configuration. The first 'rigid' component is modeled on commercially available endoscopes, typically around 12 mm in diameter and 27 mm in length. The 'soft' component facilitates the lengthening of the capsule endoscope, enabling it to adapt to the zigzag topography of the intestinal environment. This segment allows the capsule endoscope to undergo passive deformation under intestinal compression stresses and peristaltic forces, drawing on the mechanical principles inherent in traditional rotating joints and flexible surgical robots.

In this study, we present a narrowly focused investigation into the engineering design alternatives and ex vivo feasibility of the FCE. Notably absent, however, is a consideration of critical clinical parameters. Most prominently, in Figure 5, the prototype exerts a force that, at its zenith, approaches 2N—potentially sufficient to inflict mucosal injury or more grievous perforations within the delicate milieu of the small intestine. Furthermore, the current design's intermediate soft segment, critical for accommodating the undulating motions of the small intestinal tract, mandates empirical validation through in vivo experimentation to confirm its compliance and functional integrity. Our research indicates that magnetic origami structures contribute to more ergonomic and intuitive human-computer interfaces due to their flexible and adaptable nature \cite{yuan2022versatile,cai2023magnetically,yeow2022magnetically}. Meanwhile, the automated control of the FCE predicates upon real-time positional feedback. The feasibility of such a system, while preliminarily substantiated in prior investigations concerning magnetic localization techniques \cite{song2021magnetic,su2023amagposenet,su2023magnetic}, remains an avenue for prospective development.

%To enhance the capsule endoscope's traversal of the small intestine, we integrated an additional rigid section at the tail. This performs two main functions: initiating a worm-like crawling propulsion motion under the influence of an external magnetic field via the internal interaction of a pair of radial magnets, and integrating rigid functional components. Moreover, our study indicates that capsule endoscopes can expand in response to high-frequency electromagnetic heating stimulation, leading to the expansion of an Ecoflex film. This innovative attribute opens up possibilities for future capsule endoscope applications such as anchoring, drug delivery, sampling, and scaffold deployment.

\bibliographystyle{ieeetr}
\bibliography{reference}

\end{document}